\documentclass{amsart}

\usepackage{amscd}
\usepackage{amssymb}
\DeclareFontFamily{OMS}{cmsy}{
\fontdimen16\font=3pt
\fontdimen17\font=3pt}
\newenvironment{pf}{\proof[\proofname]}{\endproof}
\def\dj{d\kern-.30em\raise1.25ex\vbox{\hrule width .3em height .03em}}
\def\Dj{D\rlap{\kern-.70em\raise0.75ex
\vbox{\hrule width .3em height .03em}}}
\newtheorem{thm}{Theorem}[section]

\theoremstyle{definition}
\newtheorem{defn}{Definition}[section]
\numberwithin{equation}{section}
\def\bla#1{$(${\it #1\/{}}$)$} 
\def\id{\mathrm{id}}

\def\cal{\mathcal}
\def\End{\mathrm{End}}
\renewcommand{\thepage}{\ifnum\value{page}=1 \else\arabic{page}\fi}
\begin{document} 
\title[Generalized Noiseless Quantum Codes]{Generalized Noiseless Quantum Codes utilizing 
Quantum Enveloping 
Algebras}

\author[M. {\Dj}ur{\Dj}evich, H.E. Makaruk, R.Owczarek]{micho {\Dj}UR{\Dj}EVICH*, 
hanna e. MAKARUK**,  robert OWCZAREK**}

\thanks{\large *Instituto de Matematicas,
UNAM, Area de la Investigacion Cientifica, Circuito Exterior, Ciudad Universitaria, Mexico DF, 
CP 04510, Mexico\\[3mm]
**MS E517, E-ET, Los Alamos National Laboratory, Los Alamos, New Mexico 87545
\\[.6cm]
Short Title: {\bf Generalized Noiseless
 Quantum Codes},\\[6mm]
PACS No: 03.67.Lx, 02.20.Os, 02.20.Sv\\[10cm].
}

\maketitle

\newpage
\begin{abstract}
A generalization of the results of Rasetti and Zanardi concerning
avoiding errors in quantum computers by using states preserved by evolution
is presented. The concept of the dynamical symmetry is generalized from the level of
classical Lie algebras and groups, to
the level of a dynamical symmetry based on quantum Lie algebras and quantum groups
(in the sense of Woronowicz).
An intrinsic dependence of the concept of dynamical symmetry on the
differential calculus (which holds also 
in the classical case) is stressed. A~natural connection between quantum 
states invariant under a quantum group action, and quantum states
preserved by the dynamical evolution is discussed. 
\end{abstract}
\newpage
\section{Introduction}
\enlargethispage{\baselineskip}
Quantum computation is a new and quickly developing area of science.  Its power
comes from using quantum parallelism  of computations. This new paradigm for
computation was envisioned by Feynman \cite{Feynman}. For many years quantum computation looked as an unrealistic dream. The reason is unavoidable decoherence 
due to the interaction of quantum devices with a classical  environment, 
which destroys quantum coherent states. Then the advantage of parallelism
is lost, and this makes quantum  computation 
impossible. This situation 
changed radically when the quantum error correcting codes 
were invented \cite{Shor,Steane,Laflamme}. This fact plus remarkable 
progress in experimental manipulation with individual
qubits make the dream coming true. 

In this paper  we study a special implementation of so-called noiseless 
quantum codes, also known as error avoiding quantum codes \cite{Chinczycy}.
Such codes were proposed in \cite{Rasetti, Rasetti2}  
as an alternative or, more likely, supplement to the error correcting 
quantum codes.  In \cite{Rasetti, Rasetti2} error
avoiding quantum codes were built using group theoretic methods. 
The idea is 
that among quantum states of the system 
there exist distinguished ones which, despite interaction with the environment, do 
not underlie decoherence. Important assumption was that qubits of the 
quantum register interact with a coherent environment. This assumption, besides the assumption  on dynamical symmetry of the system, turned out to be essential for the introduction of the states protected against decoherence.  
Namely, it is possible to introduce collective variables describing the qubits composing the 
register. The singlet state of the qubits turned out to be protected against corruption. 

An attempt to describe a more general situation was made in \cite{Chuang}. There the 
semigroup technique was used to describe general evolution of the system 
interacting with environment. In comparison to 
\cite{Rasetti, Rasetti2}, the generalization consisted in  consideration of various degrees
 of coherence of the environment on the distances of the order of the 
length of the register. Besides full coherence, lack of any coherence 
and partial coherence were considered, too. Basic results on error protected 
states are similar to the ones obtained in \cite{Rasetti, Rasetti2}. 
Additional noises were considered, and it was shown that their  
influence on the error protected states is  negligible up to the 
first order of a small parameter characterizing the noise. Performing quantum computations
 with the error protected states was shown to be realistic.

In the paper \cite{Chinczycy}  general criteria for error avoiding 
quantum codes were formulated. First, the existing codes were divided 
 into three groups: error correcting codes, error avoiding codes and 
error preventing codes. Error correcting codes detect and correct errors.
 Error preventing codes only 
detect errors, without correcting them. After the classification was 
made the general theory of error avoiding codes was formulated,
in the manner
 following the paper \cite{Many}, where the general theory of error
 correcting quantum codes was presented. It was found that the error
 avoiding codes are derived from the subspaces of the Hilbert space 
that are common eigenspaces of the operators $A_a$ describing the 
evolution of the system. If $\rho_i$ and $\rho_f$ are the initial 
and the final density matrices of the system respectively, then, 
under assumption  that initially  the system  is not entagled with the environment,  
$\rho_f=\sum_aA_a\rho_iA_a^+$, where $\sum_aA_a^+A_a=I$.
 
>From the paper \cite{Many} the conditions for the error correcting
codes are known. Namely, if the vectors $|i>$ form an orthonormal basis of the 
code, the condition 
\begin{equation}
<i|A^+_aA_b|j>=\gamma_{ab}\delta_{ij}\label{1*}
\end{equation}
should be satisfied, where $\gamma_{ab}$ is a Hermitian matrix. All 
known error correcting codes have $\gamma_{ab}$  nondegenerate. The 
matrix is  then expressible, after some unitary redefinition of 
$A_a$, as a diagonal matrix with positive entries. 

As shown in 
\cite{Chinczycy} error avoiding codes also satisfy (\ref{1*}) but 
are maximally degenerate  (in the diagonal form only one diagonal 
element is different from zero). Moreover the matrix  $\gamma_{ab}$
 is expressible as  $\gamma_{ab}=\gamma_{a}^*\gamma_{b}$, where  
$\gamma_{a}$ are eigenvalues of the respective operators $A_a$ for the states 
from the code. This general approach does not use the group theoretic
 language, so that we do not know if there exist error avoiding codes 
of different origin than the group theoretic (or quantum group
theoretic) one. It shows however  
usefulness of the error avoiding codes for quantum computing (especially when used  
simultaneously with the error correcting codes). 

The aim of this paper is to study error avoiding codes in a more
general framework than the group-theoretic one. More precisely, 
we shall discuss the problematics of error avoiding codes in the framework
of quantum groups. 

Our motivation is that the quantum group framework enables 
introduction of a more general dynamical symmetry of the system, comparing to
the standard group theoretic one. By construction, it covers also  the dynamical
symmetry connected with classical groups,  
since  the  classical groups  are all special cases of
quantum groups. We hope to take this way into account
certain {\it fluctuations} from the exact group theoretic dynamical symmetries 
required by the standard noiseless codes.

In \cite{Rasetti, Rasetti2} there was  expressed 
a hope that deviations from the proposed ideal situation should 
not destroy too much of the quantum coherence and consequently 
the error protected states should become 
`almost error protected' under these conditions. Our considerations show 
there exist particular perturbations
for which the error protected states `remain' {\it exactly} protected despite 
these perturbations.

\medskip
The plan of the paper is as follows. In the second section we define
 and briefly discuss the notion of dynamical symmetry connected with
 quantum groups. In the third section  we formulate and prove main 
theorems concerning error protected states. In the fourth section 
we present simple examples  to illustrate the general results. 
We conclude the paper with the discussion  of possible extensions
 of this work.  In the Appendix we give a very brief review of the basic 
material on quantum groups and their representations. 
 
\section{Dynamical Symmetry Coming From Quantum Groups}
 
Symmetry proved to be one of the basic notions in physics. Dynamical
symmetry of a physical system 
is defined in terms of its Hamiltonian, which should be expressible
as a  linear 
combination of operators generating a representation of the appropriate
Lie algebra.
There is a large class of systems possessing such a property. Dynamical 
symmetry of a system should not necessary  be  visible at a first sight. 
Nevertheless, searching for such a symmetry is highly rewarding,
since one can apply to
the systems with a dynamical symmetry powerful methods 
developed on the ground of the theory of Lie algebras and
their representations, like the method of coherent states 
\cite{Pierelomov}.
Dynamical symmetry proved to be also important  in searching 
for physical systems possessing very specific 
quantum states---which can not 
be corrupted despite their interaction with the 
environment \cite{Rasetti,Rasetti2}. Such
systems provide noiseless quantum  codes that are of potential 
great interest for constructing quantum computers.  Noiseless 
quantum   codes can be either alternative to
error correcting codes, which are elaborate methods of
coding information, recognizing 
errors and correcting them \cite{Shor,Steane,Laflamme}, 
or a valuable supplement to such codes.

It turns out that analogous codes exist for systems 
with dynamical symmetry based on quantum groups instead of
Lie groups.
The goal of this section  is to define the notion of
 dynamical symmetry 
associated to quantum groups. In the next section we 
shall apply our new concept of dynamical symmetry
to prove our main theorems concerning error protected states. 
Then, we shall study  some systems providing noiseless quantum codes. 

Basic mathematical concepts and tools that will be used in the paper are briefly
presented in the Appendix. 
 
A generalization of the concept of dynamical symmetry can
be defined only when there are well established notions 
of a Lie algebra, and the corresponding universal enveloping algebra, 
associated to a given quantum group $G$.  In the theory of quantum groups, 
all these notions essentially depend on an appropriately chosen 
{\it differential calculus} over $G$. 

The quantum group $G$ will be represented by 
a non-commutative $C^*$-algebra $A$, playing the role of the algebra of `continuous 
functions' defined on the quantum space $G$, together with a coproduct 
map $\phi:A\rightarrow A\otimes A$ (corresponding to the standard product in 
the case of classical groups). Effectively, all caclulations will be performed 
within an everywhere dense *-subalgebra $\mathcal{A}\subseteq A$, 
playing the role of polynomial functions over $G$. Actually, $\mathcal{A}$ is
a Hopf *-algebra in a natural way. 

Suppose that on $G$ is defined a *-covariant, left-covariant first-order
differential calculus $\Gamma$. Let $L$ be the associated quantum Lie algebra. 
If the calculus $\Gamma$ is in addition right-covariant, we can introduce the universal enveloping algebra
$U(L)$. Every representation $v: V\rightarrow V\otimes \mathcal{A}$ of $G$ in a finite-dimensional 
vector space $V$ 
naturally induces (as in the classical theory) 
a representation $\delta=\delta_v$ of $L$ and $U(L)$ in $V$. 

Definitions of all these objects are  sketched in the Appendix.

We consider an open quantum system, represented by
a Hilbert state space $V=H_S$. 
The system interacts with its environment (bath) which is described by a Hilbert 
space $H_B$. Here it is assumed for simplicity that all Hilbert 
state spaces are finite dimensional---however,  everything could 
be incorporated into the infinite-dimensional case.    

\begin{defn}
We say that a system has quantum  dynamical 
symmetry  described by the quantum  group $G$ 
and its quantum  Lie algebra $L$ 
if the following conditions are 
satisfied: 

\smallskip
\bla{i} The evolution of the system is 
governed by the Hamiltonian 
$$\mathcal{H}\in \mbox{\rm End}(H_S \otimes H_B)\simeq 
\mbox{\rm End}(H_S)\otimes \mbox{\rm End}(H_B).$$

\smallskip
\bla{ii} The Hamiltonian is a hermitian operator $\mathcal{H}^*=\mathcal{H}$. 

\smallskip
\bla{iii} The Hamiltonian is of the form:
\begin{equation}
\mathcal{H}=P_1(l_1,\ldots,l_n)\otimes T_1+\ldots +P_N(l_1,\ldots, l_n)\otimes T_N
\label{3}\end{equation}
where $P_1,\ldots ,P_N$ are polynomial expressions 
of infinitesimal generators $l_i=\delta(e_i)$ and $\{e_i\}$ 
is a basis in $L$. Finally $T_1,\ldots,T_N$ are hermitian operators
$$T_{\alpha}\colon H_B\rightarrow H_B.$$
\end{defn}

Such systems with quantum  dynamical symmetry can  
be explored by generalized methods known  from 
the theory of systems with classical dynamical symmetry, for example by the 
method  of quantum coherent states \cite{Demichev}.
Let us observe that the terms in (\ref{3}) can be reorganized in such a way,
that the Hamiltonian takes a more familiar form:
\begin{equation}
\mathcal{H}=\mathcal{H}_S\otimes \id+\id\otimes \mathcal{H}_B+\mathcal{H}_I
\label{4}\end{equation}
where $\mathcal{H}_S$ is the system's Hamiltonian, 
$\mathcal{H}_B$ is the Hamiltonian of the environment 
and $\mathcal{H}_I$ is the `interaction hamiltonian' uniquely defined as the part of $\mathcal{H}$ traceless 
in both tensor factors. 

\section{Error Protected States $\&$ Noiseless Quantum Codes}

In this section we present our main theorems on error protected
states, and on noiseless quantum  codes. 
We assume that  we deal with a (open) 
quantum system with dynamical symmetry of a quantum group, 
and all other features  as described in  the previous section.
The vectors that are $v$-invariant, where $v$ is a representation
of the quantum group $G$, are of vital importance for our further
discussion. Let us give their definition now. 
Let $v:V\to V\otimes {\cal A}$ 
be an arbitrary representation of 
$G$ in a finite-dimensional vector space $V$, 
and let $\delta=\delta_v:L\to \mbox{\rm End}(V)$
be the associated representation of $L$.
To further simplify the considerations, 
we shall consider the case when the 
quantum group is `connected' in the sense that only scalar elements of
$\mathcal{A}$ are annihilated by the differential $d\colon\mathcal{A}\rightarrow\Gamma$.

Then  the following equivalence holds
for every vector $\psi\in V$

$$
v(\psi)=\psi\otimes 1\quad\Leftrightarrow\quad
\delta(x)\psi=0, \quad\forall x\in L
$$

Let us assume that the calculus $\Gamma$ is in addition bicovariant. 
This enables us to introduce the quantum universal enveloping algebra $U(L)$, and to discuss the 
representations of $U(L)$ associated with
the representations of the quantum group $G$. Let us also 
introduce the map $\chi:U(L)\to \mathbb{C}$, 
with the properties $\chi(L)=0$, $\chi(1)=1$, extended then  to the whole
$U(L)$ by multiplicativity. 
The representation $\delta$ uniquely (as in 
the standard theory) extends from $L$ to $U(L)$. The above two 
conditions are further equivalent to
$$\delta(q)\psi=\chi(q)\psi, \quad\forall q\in U(L).$$ 

The proof of these equivalences is quite straightforward, but
it needs some additional definitions and constructions,
which we would rather skip in this paper as they are going too far
in the formalism. 
Vectors satisfying  any of the above 
conditions are called {\it v-invariant}. 
The $v$-invariant vectors are very important for the study of quantum
registers (which are open systems with a quantum dynamical symmetry). 
Such vectors give us examples of the error protected states.

Our main theorem reads:

\begin{thm}\label{thm1}
Unitary evolution  described by the Hamiltonian $\mathcal{H}$ possessing a 
quantum dynamical symmetry given by $(G,L)$
preserves the $v$-invariant vectors and associated states of the 
system,  even when  the other states of the system are 
corrupted due to decoherence.
\end{thm}
\begin{pf}
Let us take  as an initial vector 
$\psi\otimes \zeta\in H_S\otimes 
H_B$, where $\psi$ is $v$-preserved 
in the sense defined above.
Then the unitary evolution defined by 
$$U(t)=\exp(-\frac{i}{\hbar}\mathcal{H}t)$$ 
with $\mathcal{H}$ of the form (\ref{4})
gives
$$
\exp(-\frac{i}{\hbar}\mathcal{H}t)(\psi\otimes \zeta)=\psi\otimes 
\exp(-\frac{i}{\hbar}\mathcal{H}_{\mathrm{eff}}t)\zeta
$$
where
$$
\mathcal{H}_{\mathrm{eff}}=\chi(P_1)T_1+\ldots +\chi(P_N)T_N.
$$
This proves the statement. 
\end{pf}

Interesting property of $\mathcal{H}_{\mathrm{eff}}$ 
is that  the coefficients  $\chi(P_i)$ should  somehow
reflect the structure given by $G$ and its Lie algebra $L$.

Now we can easily prove generalization of the 
theorems 1 and 2 given in the paper \cite{Rasetti}.
We follow the notation from \cite{Rasetti}. Let 
$\rho_S\in \mbox{\rm End}(H_S)$ and $\rho_B\in \mbox{\rm End}(H_B)$ be the (mixed quantum)
states of the system and the environment respectively. If the overall 
system is initially in the state $\rho(0)=\rho_S\otimes \rho_B$, 
then $\rho(t)=U(t)\rho(0)U(t)^+$, so that the evolution is unitary.
The induced evolution on $H_S$ is given exactly like in \cite{Rasetti} by 
$L_t:\rho_S\to \mbox{\rm tr}_B \rho(t)$,
where $\mbox{\rm tr}_B$ is the trace over $H_B$. Then the following theorem is 
fulfilled:

\begin{thm}\label{thm2}
Let ${\cal M}_N$ be the manifold of 
states built over the space  of vectors 
invariant under the 
representation $v$, and 
$\rho_S\in{\cal M}_N$.
Then for any initial bath state $\rho_B$ 
the induced evolution on $H_S$ is trivial, 
$$L_t[\rho_S(t)]=\rho_S, \qquad \forall t>0.$$
\end{thm}

\begin{pf}
Theorem \ref{thm1} allows us to reduce the proof of \ref{thm2} to the proof of the first
theorem of \cite{Rasetti}. 
\end{pf}

The invariant vectors are generalizations 
of the singlet states pointed out in \cite{Rasetti} as the states of the 
quantum register which  are not corrupted  by interaction with the environment.

We should stress that the Hamiltonian of the system + environment should 
not necessary contain terms with trivial representation in the 
space of the system and in the space of the environment, so 
that it can be even of the pure interaction form.
 
Before we present simple examples illustrating the general theory  
and explicitly showing the `error protected' states, let 
us discuss interesting question of the structure of the Hilbert
space of the quantum computer registers, and discuss
physical implications.  The register usually consists 
of a number of copies of the same quantum system, often  
having two possible states for example spin `up' and spin `down' 
(a qubit).

Dynamical symmetry is defined in the Hilbert space that originates from the Hilbert
spaces for individual qubits being described as carrier spaces of unitary 
representations 
$$v_i:V_i\to V_i\otimes{\cal A}\qquad i=1,\ldots, n$$
of our quantum  group $G$. 

The register Hilbert space is the tensor product of the 
representation spaces,  
$$V=V_1\otimes V_2\otimes\ldots\otimes V_n$$
in which $G$ acts by the direct product 
$$v=v_1\times v_2\times \ldots \times v_n$$
of representations $v_i$. Each of the representations 
$v_i$ induces a representation $\delta_i$ of the corresponding quantum 
universal enveloping  algebra. The representation  $v$ induces the 
representation $\delta$ of the  quantum universal enveloping algebra, and one can
easily prove the following relation:
\begin{equation}
\delta(x)(\phi_1\otimes\ldots\otimes \phi_n)=\sum_{k=1}^n \sum_{\alpha\in I[k]}
\phi_1\otimes\ldots\otimes\delta_k(x^\alpha)\phi_k\otimes\eta_{k+1}^\alpha\otimes\ldots
\otimes\eta_n^\alpha
\label{5*}\end{equation}
where 
$$\tau_{n-k}\Bigl(\Bigl\{\phi_{k+1}\otimes\ldots\phi_n\Bigr\}\otimes x\Bigr)=
\sum_{\alpha\in I[k]} x^\alpha\otimes\Bigl\{\eta_{k+1}^\alpha\otimes\ldots\otimes
\eta_n^\alpha\Bigr\}$$ and 
$\tau_{n-k}: V_{k+1}\otimes\ldots\otimes V_n\otimes L\to L\otimes V_{k+1}\otimes\ldots\otimes V_n$
are the appropriate `flip-over' operators naturally associated to 
the differential calculus. 

The formula (\ref{5*}) differs from the corresponding formula for the classical
case of addition angular momenta in quantum mechanics ($\tau$
is in the classical case just the standard 
transposition).  It is easy to see that qubits in the register are not treated on the same footing.  It
could be associated to some effects due to,  not taken into account in
\cite{Rasetti}, linear  extension of the register, or to fluctuations  
of the fields due to nonideal structure of boundaries of the register and their 
influence. Anyway, it is possible to realize a system with weaker symmetry 
than the one presented in \cite{Rasetti} but still possessing error protected states. 
It is known that similar
deviations from exact dynamical symmetry of Lie groups lead to 
better mass (or energy) formulas in both nuclear/particle physics, and
molecular physics \cite{Iwao,Gavrilik}.
Therefore, one can look also among such systems for possible candidates for
registers of quantum computers.

In \cite{Rasetti} a physically plausible conjecture was expressed, that small 
deviations from ideal properties assumed of the 
system should lead to small errors  in the error-protected 
states. Actually, we have shown that there exist systems with 
special kind of deviation from the assumed symmetry, which
nevertheless still have error protected states.

\section{Examples}
Let us switch now to some simple examples that would 
highlight our general ideas. The first
example of a quantum group presented systematically in 
the literature was a quantum version of the standard $SU(2)$ 
group \cite{RIMS}, where the theory of representations was developed 
together with various geometrical aspects and a construction of 
a natural three-dimensional left-covariant differential calculus. 
This calculus is not bicovariant, and the minimal
dimension for a bicovariant calculus over the quantum $SU(2)$ is 4 (this
four-dimensional calculus \cite{Woronowicz2} is analyzed in detail in 
\cite{Stach}). Generalization of the results concerning this particular quantum 
group leads to the general theory of compact matrix quantum groups 
\cite{Woronowicz1,Woronowicz2}, the definition of quantum spheres
\cite{Podles1} and their geometry \cite{Podles2}, deep generalization of the
Tannaka-Krein duality 
\cite{Woronowicz3}, and also the theory of quantum principal 
bundles together with the corresponding gauge theory on quantum spaces, 
first formulated in \cite{Micho1} and then developed 
systematically in \cite{Micho2, Micho3} (see also \cite{Micho4, Micho5, Micho6}). 
Also in the $C^*$-algebraic 
framework the quantum homogeneous 
bundles were defined and the example 
of such a bundle with quantum spheres
as fibers was given \cite{Hania}.
Different approaches to  quantum groups were developed in
\cite{Drinfel'd, Fad-Resh-Takh}, where quantum groups are 
treated from the point of view of deformations of
universal enveloping algebras, Yang-Baxter equations and
completely solvable systems.

We shall use the quantum version of
$SU(2)$ in our examples. This is relatively simple from computational
viewpoint, but highly non-trivial and
very suggestive for the aims of this paper. First, we remind 
some basic facts about this group, which is denoted by $S_{\mu}U(2)$. 
Here the deformation parameter $\mu$ takes the 
values $\mu \in [-1,1]\setminus \{0\}$, and $\mu=1$
corresponds to the classical $SU(2)$ group.

In our further considerations important role will play the fact 
that irreducible unitary representations of $S_{\mu}U(2)$ are classified by  
the half-integers, like the representations of $SU(2)$. 
The fundamental representation corresponds, as in the classical case,  to
spin $j=\frac{1}{2}$ (see Appendix for more details). The Clebsch-Gordan decompositions
of tensor products of the representations of the $S_{\mu}U(2)$ into irreducible 
representations look similar 
(concerning the multiplicities of the appearance of irreducible
components in the products of representations) as in the classical case:
$$
\overbrace{u\times \dots\times u}^k=\bigoplus_{j \in{J}}n_{j,k}u_j
$$
with the numbers $n_{j,k}$ the same as in the classical case. 
In particular, the decomposition 
of the second tensor power of the 
fundamental representation is 
$u_{1/2}^{2}=u_0 \oplus u_1$,
 where $u_0$ and $u_1$ are the 
$1$-dimensional and the $3$-dimensional irreducible
 representations, respectively. 
One can describe these representations more explicitly after introducing 
the orthonormal basis in the representation space 
$V=\mathbb{C}^2$ of $u_{1/2}$, which will be denoted $\vert+\rangle$, 
$\vert-\rangle$ for the purpose of being easily recognizable 
by physicists. The tensor square  $u_{1/2} ^{2}$
 is realized in ${V \otimes V }\simeq \mathbb{C}^4$, and the orthonormal basis in this space is $\vert+
\rangle\otimes\vert+\rangle$, $\vert+\rangle\otimes\vert-\rangle$, 
$\vert-\rangle\otimes\vert+\rangle$, $\vert-\rangle\otimes\vert-\rangle$. 
It is an easy exercise to find that the invariant
 subspaces of $u_{1/2}^{2}$ are spanned by:
\begin{equation}
\frac{1}{\sqrt{1+{\mu}^2}}
(\vert +
\rangle
\otimes
\vert
 -\rangle 
-\mu\vert -
\rangle\otimes\vert 
+\rangle)
\label{singlet}
\end{equation}
and 
\begin{equation}\label{triplet}
\begin{gathered}
\vert +\rangle\otimes\vert +\rangle\\
\frac{\mu}{\sqrt {1+{\mu}^2}} (\vert +\rangle\otimes\vert
 -\rangle +\frac{1}{\mu} \vert -\rangle\otimes\vert +\rangle )\\
\vert -\rangle\otimes\vert -\rangle
\end{gathered}
\end{equation}

The formula (\ref{singlet}) generalizes the standard singlet, and 
the formula (\ref{triplet}) generalizes the standard triplet. 
In analogy to the classical case 
the even tensor powers of the fundamental representation 
decompose into irreducible 
representations in such a 
way that the one-dimensional representation appears a number 
of times, and the number is identical as in the classical 
case. These singlets will be preserved by the dynamics.

\subsection{First Example}

In the first example we treat a system which is very similar to the 
one considered in \cite{Rasetti}. Namely, as a model of the
 environment (bath) we 
consider a system of harmonic 
oscillators, described by the
 Hamiltonian 
$$\mathcal{H}_B = \sum_{k} \omega_{k}b_{k}^{\dagger}b_{k},$$
acting in the Hilbert space $H_B$.
The register consists in this simplest case of 
two qubits. In contrast to the case considered
by Zanardi and Rasetti \cite{Rasetti}, the system consisting of the 
register and the bath has the dynamical symmetry 
not of the classical but of the quantum $SU(2)$ 
group. As already mentioned, 
in the quantum group
context it is necessary to chose the 
differential calculus, prior to establishing the notion of
the dynamical symmetry. The closest to the classical case seems to be 
introduction of the $3D$ left-covariant calculus \cite{RIMS}. In other words, the 
quantum Lie algebra $L$ is $3$-dimensional. 
Let us denote by $K_i$ the operators 
representing the basis vectors $l_i$, in an arbitrary 
representation of $L$ (here $i\in\{1,2,3\}$). 
The following recurrent formulas enable
us to compute explicitly the operators $K_i$, 
in the arbitrary tensor product of elementary
$2$-dimensional representations (qubits):
\begin{eqnarray}
K_3(\psi\otimes\vert{+}\rangle)&=\frac{1}{2}\psi
\otimes\vert{+}\rangle+\frac{1}{\mu^2}
K_3(\psi)\otimes\vert{+}\rangle\\
K_3(\psi\otimes\vert{-}\rangle)&=\mu^2
K_3(\psi)\otimes\vert{-}\rangle-\frac{1}{2}\psi\otimes\vert{-}\rangle\\
K_j(\psi\otimes\vert{+}\rangle)&=\frac{1}{2}\psi\otimes\vert{+}
\rangle+\frac{1}{\mu}
K_j(\psi)\otimes\vert{+}\rangle\\
K_j(\psi\otimes\vert{-}\rangle)&=\mu K_j(\psi)\otimes\vert{-}\rangle
-\frac{1}{2}\psi\otimes\vert{-}\rangle,
\label{reps}\end{eqnarray}
where $j\in\{1,2\}$. 

In such a case the bath-register interaction Hamiltonian which
is the quantum group analog of the Hamiltonian used in \cite{Rasetti} reads:
$$
\mathcal{H}_I=K_+\otimes T+K_{-}\otimes T^\dagger+K_3\otimes T',
$$
where $K_{\pm}=K_1\pm iK_2$, and $T,T'$ are operators acting in the bath
Hilbert state-space. Relating to the corresponding formulas in \cite{Rasetti},
the operators $T$ and $T'$ are obtained as the appropriate linear 
combinations of the creation and annihilation operators describing 
relevant elementary excitations of the bath.
The operators $K_j$ are acting in the 
$4$-dimensional 2-qubit space. 
In other words, it is formally of the same shape as in \cite{Rasetti},
however the
`spin' operators are different as explained above. The 
singlet state of the register is error-protected in the 
sense discussed above. 

\subsection{Second Example}
In the second example the only difference with the first example is the 
register consists now of 
any even number of qubits,
instead of just two. The spin operators $K_j$ are referring to the total
register system, and are calculated by applying the above listed
inductive rules (\ref{reps}). 

It is important to stress that the number of singlet states 
is just the same as in the 
classical $SU(2)$ case. This is a consequence of the mentioned similarity between
the representation theories for quantum and classical $SU(2)$ groups.
The dimension of the singlet state space depends on the number of 
qubits in the way described 
in \cite{Rasetti}. All these
states are clearly protected
from corruption due to decoherence. 

\subsection{Remark}
The group $SU(2)$ appears  as a dynamical symmetry group mainly in the  context 
of the dynamics of spin  systems. It appears less frequently  in the context of 
dynamics of different systems, like bosonic particles.  Therefore,  the examples  
usually begin  with  the fundamental representation $u_{1/2}$ of 
$SU(2)$, as the elementary building blocks of the Hilbert space  of the system. 
However, such a fundamental block could be any of 
$u_j$ with $j$ half-integer. Since for example $u_j\times u_j$ 
contains the singlet $u_0$ in its splitting into  irreducible representations.
this could be a starting point
for building error  protected states  with the  help of   bosonic particles. 
Similar considerations are true also  for the quantum group 
$S_\mu U(2)$. It seems, though, that a physical realization  of such systems  is
more complicated and creates more technical problems.
 
Let us stress that in the examples we
considered following the paper \cite{Rasetti}, all qubits are coupled to the
same, coherent, environment. As was stressed in \cite{Chinese} coupling to the
same environment of the qubits gives more possibility to get error protected
states than coupling to independent environments. However, our methods
seem to be general enough to deal with the cases of coupling with
independent environments as well, till the system has a dynamical symmetry of
the type introduced in this paper.

\section{Conclusions}
In this paper we introduced the general notion of 
dynamical symmetry associated with quantum groups and Lie algebras. 
Then we applied this notion to construct error
protected states for open systems with such symmetry.
The states can be  useful  for quantum  computation. They  are close analogs of their 
standard group theoretical counterparts. As a result, the error protected  states  obtained 
in  strictly group theoretical dynamical symmetry  context have counterparts  preserved when the 
symmetry  is slightly deviated towards  the quantum  group theoretic one. 
Recently various authors \cite{Lloyd,Lloyd2} \cite{KnillLloyd} \cite{Chinczycy3} \cite{Vitali} 
introduced  a technique of
quantum computation which
dynamically eliminates  errors, by a quantum counterpart of the classical so
called `bang-bang' technique. Zanardi \cite{Zanardi} has shown that the technique 
called `symmetrizing' can be interpreted  group theoretically as control
of the systems forcing the  systems with dynamical symmetry  of a Lie group to be
in states which are error protected. This very interesting observation  
should increase interest in error avoiding quantum codes.
Since mathematically the technique  seems 
to rely  on invariant measures on the groups,
 it is applicable  not only  to the systems with the dynamical symmetry of 
 finite groups  discussed  
 in the paper, but also to the systems with dynamical symmetry 
 of the compact (even locally compact) Lie groups,  which all possess Haar measure necessary for
such construction.  One should observe that the same  is also true for compact (or locally compact) 
quantum groups, since these 
objects possess the Haar measure, too. It seems the generalization of the results by Zanardi  
to the quantum group case is straightforward, but its physical interpretations
are less clear. Work on this issue is in progress. 

\section*{Acknowledgements}M. {\Dj}ur{\dj}evich likes to thank  
Theoretical Division of Los Alamos National Laboratory for the hospitality, 
H.E. Makaruk and R. Owczarek like to 
acknowledge the hospitality of
the Math Institute of National Autonomous University of Mexico. 
Realization of this research was partially supported by 
Investigation Project in106879 of DGAPA/UNAM

\appendix
\section{Quantum Spaces and Quantum Groups}

Classical theorem by Gelfand and Naimark states that
compact topological spaces are in a natural correspondence with commutative unital
$C^*$-algebras. These $C^*$-algebras consist of
continuous complex-valued functions on the corresponding spaces.

Let $X$ be a compact topological space and $C(X)$ be the associated algebra of 
continuous complex-valued functions on $X$.
The linear structure on $C(X)$ is given by the obvious 
conditions:
$(f+g)(x)=f(x)+g(x)$, and $(\alpha f)(x)=\alpha f(x)$. The product in
the algebra is $(f\cdot g)(x)=f(x)g(x)$ and the *-involution
is given by $f^{*}(x)=\overline{f(x)}$.

There is a natural norm in $C(X)$ given by
$$||f||=\sup\{ |f(x)|:x\in X \}.$$

In such a way is introduced a structure of (commutative) $C^*$-algebra in $C(X)$. 
Conversely, every commutative unital $C^*$-algebra is of this form---according to
classical Gelfand-Naimark theory.  Actually, the Gelfand-Naimark theory can be generalized 
to the level of {\it locally-compact} spaces, giving us a correspondence between
arbitrary commutative (not necessarily unital) $C^*$-algebras and locally compact
topological spaces. This correspondence is
functorial, in the sense of category theory. 

These facts lead to a generalized concept
of space, which is understood as `the underlying space' of a general $C^*$-algebra, 
about which we no longer assume it should be commutative. 
Generalized spaces of this type are called quantum spaces. 
The reason for the adjective `quantum'
follows from the observation that as in the classical
quantization scheme a commutative algebra of functions is substituted
by a noncommutative algebra of operators acting in a Hilbert space.
The latter is indeed the case since all $C^*$-algebras can be
realized as algebras of operators acting in some Hilbert spaces.

Interesting algebra and geometry appears when the classical topological
spaces are equipped with an additional structure: differential-geometric, metric,
Lie group, and so on. A very important class of quantum spaces constitute
the quantum groups, which are understandable as quantum spaces equipped with a 
group structure. 

Let us explain now, very briefly, what is exactly a compact quantum group. 
Let us start from a classical compact topological group $G$. This means that $G$ is a compact 
topological space equiped with a group structure, such that the
product map $\circ:G\times G\rightarrow G$ is
continuous (it can be shown that in the compact case continuity of the product implies 
continuity of the inverse map). At the dual level, the product map is represented by a
*-homomorphism $\phi:A\rightarrow A\otimes A$, where $A=C(G)$. 

More precisely, we first naturally identify 
$$
\overbrace{A\otimes\dots\otimes A}^k=C(\overbrace{G\times\dots\times G}^k)\qquad k\geq 2
$$
and define
$$
\phi (f)(g_1,g_2)=f(g_1g_2), \qquad f\in A\qquad g_1,g_2 \in G. 
$$

The associativity property
of the product is equivalent to the coassociativity property
$$
(\phi\otimes \id)\phi=(\id\otimes\phi)\phi.
$$ 
It can be shown that the 
remaining two group axioms (the existence of the neutral element and
the existence of the inverse elements) are equivalent to a single assumption 
that the elements of the form $a\phi(b)$ as well as of the form $\phi(b)a$, 
where $a,b\in A$, span two everywhere dense linear subspaces of $A\otimes A$. 

Generalizing this to the quantum level, we define a group structure
on a {\it quantum space} $G$ as a *-homomorhism $\phi:A\rightarrow A\otimes A$ 
such that the diagram 
\begin{equation*}
\begin{CD}
A @>{\mbox{$\phi$}}>> A\otimes A\\
@V{\mbox{$\phi$}}VV @VV{\mbox{$\id\otimes \phi$}}V\\
A\otimes A @>>{\mbox{$\phi\otimes \id$}}>
A\otimes A\otimes A
\end{CD}
\end{equation*} 
is commutative, and such that
\begin{align*}
A\otimes A&=\overline{\Bigl\{\sum a\phi(b)\vert a,b\in A\Bigr\}}\\
A\otimes A&=\overline{\Bigl\{\sum \phi(b)a\vert a,b\in A\Bigr\}}.
\end{align*}
where the bar means appropriate closure.

As a very important special case of this structure, 
let us mention {\it matrix} groups. These structures
are given by triplets $(A,\phi,u)$ consisting of a $C^*$-algebra $A$, a *-homomorphism 
$\phi\colon A\rightarrow A\otimes A$ and a matrix $u\in M_n(A)$ (all $n\times n$-matrices with
coefficients from  $A$) which is (together with the conjugate matrix $\bar{u}$)
invertible in $M_n(A)$ and such that

\smallskip
\bla{i}
The *-algebra $\mathcal{A}$ generated by the entries $u_{ij}$ 
is everywhere dense in $A$; 

\bla{ii}
The following identity holds:
$$
\phi(u_{ij})=\sum_k u_{ik}\otimes u_{kj}. 
$$

In this case we have the inclusion
$$\phi(\mathcal{A})\subseteq\mathcal{A}\otimes_{\mathrm{alg}}\mathcal{A}.$$
Let us stress that the above mentioned coassociativity and density properties are 
satisfied automatically. 

Matrix groups generalize compact Lie groups (if $A$ is commutative the theory
reduces to standard compact matrix groups). 

The algebra $\mathcal{A}$ plays the role of the algebra of 
polynomial functions over $G$. The matrix
$u\in M_n(A)$ correspond to the fundamental representation of the group $G$. 

\section{Differential calculus on quantum groups,
Quantum Lie Algebras, Quantum universal envelopes}

\subsection{Quantum Lie Algebras}

There is a very important notion of a differential 
structure defined for quantum groups. The definitions of 
a quantum Lie algebra and of a quantum universal enveloping
algebra depend on the introduced differential calculus.
Therefore, we begin from giving the definition of the
differential calculus.

First-order differential calculi are defined as 
certain {\it bimodules} $\Gamma$ 
over ${\cal A}$, equipped with 
a {\it differential} $d\colon{\cal A}\rightarrow \Gamma$. 
The space $\Gamma$ is a noncommutative counterpart of the 
usual module of $1$-forms over a classical group, 
and $d$ generalizes the standard differential of functions.

It is important to mention that there is not a unique 
prescription to construct a differential
calculus over a quantum group, and generally a given 
quantum group will possess a variety
of non-equivalent calculi, each of them having a 
potential significance. It is surprising that
the same situation appears in {\it classical theory}, 
where one can also use the methods of
quantum groups to construct new differential calculi over the {\it standard Lie groups}. This opens
interesting new possibilities in the study of classical Lie groups. In particular, it opens a 
possibility to extend the notion of dynamical symmetry, in the framework of classical groups. 

In the quantum group theory a special role is played by so-called {\it left-covariant} and
{\it bicovariant} differential calculi. In these cases \cite{Woronowicz3} we can introduce the analogs of left and
left/right actions of the group $G$ on $\Gamma$. 
If the module $\Gamma$ is left-covariant, then we can define its subspace 
$\Gamma_{\mbox{\scriptsize\rm inv}}$, consisting of
left-invariant `$1$-forms'.  Quantum Lie algebra is then defined as the corresponding
dual space, in other words $L=\Gamma^*_{\mbox{\scriptsize\rm inv}}$. 

If the calculus is bicovariant, then we can introduce a natural {\it braid operator}
$\sigma\colon L\otimes L\rightarrow L\otimes L$, generalizing the classical
transposition. Furthermore, 
in analogy with classical theory, we can 
define a quantum Lie bracket in the space $L$ generalizing the classical 
Lie bracket \cite{Woronowicz3}.
The Lie bracket is defined as a linear operator 
$C: L\otimes L\to L$, and we can equivalently write 
$[x,y]=C(x\otimes y)$. This bracket satisfies the appropriate 
generalized Jacobi identity and braided-antisymmetricity conditions. 

Following the classical theory, 
the quantum universal enveloping algebra 
for $(L,[,])$ is defined as a unital associative 
algebra $U(L)$  generated by relations 
\begin{equation}
xy-\sum_iy_ix_i=[x,y],\label{5}
\end{equation}
where $x,y\in L$ and $\sum_iy_i\otimes x_i=\sigma(x\otimes y)$.

\subsection{Representations of Quantum Groups 
and Quantum Lie Algebras}

Having the Lie bracket and using (\ref{5}) one can define
representations of quantum  Lie algebras and of the 
corresponding quantum  universal enveloping algebras.
 It can be shown that every representation $v$ of $G$ 
in a  finite-dimensional vector space $V$ naturally 
gives rise to a representation 
$S:U(L)\to\mbox{\rm End}(V)$ of  the quantum universal 
enveloping algebra.  
Namely,  let $v:V\to V\otimes {\cal A}$ be a (left)
 representation of the quantum group 
$G$ in a  finite dimensional 
complex vector space $V$, in other words $v$ is 
linear, satisfies the condition
$$(\id \otimes\phi)v=(v\otimes \id)v$$
and $v$ is invertible, understood as an element of $\End(V)\otimes \mathcal{A}$.  
This corresponds
to the classical requirements on representations of groups saying that
 the product of group elements is represented by
composition  of operators representing these 
elements, and the neutral element of a group is 
represented by the identity operator. 

Every representation $v$ of $G$ in $V$ naturally generates a representation 
$$\delta=\delta_v: U(L)\to \mbox {\rm End}(V)$$ of $U(L)$ in $V$ 
(if the differential  calculus is bicovariant)
or only of the Lie algebra $L$, $\delta: L\to \mbox {\rm End}(V)$
(if the differential calculus is left-covariant).

Moreover, if the differential calculus is *-covariant,
which  means that in the module $\Gamma$ of 1-forms 
is defined the *-operation  $^*:\Gamma\to\Gamma$ 
induced by $*$ in ${\cal A}$, it makes sense to 
speak about hermiticity of the representation $\delta$.  
Namely, the $^*$-operation  on $\Gamma$ naturally 
induces the $^*$-structure on the quantum Lie 
algebra $L$, via the formula $<f^*,\psi>=-<f,\psi^*>$ 
where $f\in L=\Gamma^*_{\mbox{\scriptsize \rm inv}}$ 
and $\psi\in \Gamma_{\mbox{\scriptsize \rm inv}}$. 

\subsection{Quantum $SU(2)$ group}

This quantum group is based on 
a $C^*$-algebra $A$
generated by elements $\{\alpha,\alpha^*,
\gamma,\gamma^*\}$ satisfying the 
following relations:
\begin{gather*}
\alpha\alpha^* + \mu^2\gamma^*\gamma=1\qquad\alpha^*\alpha + \gamma^*\gamma=
1\\
\gamma^*\gamma=\gamma\gamma^*\\
\alpha\gamma=\mu\gamma\alpha\qquad
\alpha\gamma^*=\mu\gamma^*\alpha,
\end{gather*}
where $\mu\in[-1,1]\setminus\{0\}$. The comultiplication $\phi\colon A\rightarrow A\otimes A$
is given by
\begin{gather*}
\phi(\alpha)=\alpha\otimes\alpha-\mu\gamma^*\otimes\gamma\qquad
\phi(\alpha^*)=\alpha^*\otimes\alpha^*-\mu\gamma\otimes\gamma^*\\
\phi(\gamma)=\gamma\otimes\alpha+\alpha^*\otimes\gamma
\qquad
\phi(\gamma^*)=\gamma^*\otimes\alpha^*+\alpha\otimes\gamma^*
\end{gather*}

The theory of representations of $S_{\mu}U(2)$ is 
very interesting from the point of view of our examples. 
This theory has many similarities to its classical counterpart--the theory of
representations of 
$SU(2)$. Classical $SU(2)$ is obtained as a special case $\mu=1$. 

The {\it fundamental
representation} of $S_{\mu}U(2)$ is defined by the matrix
$$
u=\begin{pmatrix}\alpha&-\mu\gamma^*\\ \gamma&\alpha^*\end{pmatrix}.
$$
It is easy to see that the defining 
relations for $S_{\mu}U(2)$ are equivalent to the unitarity property
$$
u^*u=uu^*=\begin{pmatrix}1& 0\\ 0& 1\end{pmatrix}.
$$

The fundamental representation enables us to build all other representations 
by using direct sums, tensor products and reduction procedures. Irreducible
representations $u_j$ are numbered by half-integers $j$,  and
are $2j+1$-dimensional. Every representation  of an arbitrary compact quantum group can be 
decomposed into  irreducible ones.

\end{document}